\documentstyle[prl,aps]{revtex}
\draft
\begin{document}
\input{epsf}
\twocolumn[\hsize\textwidth\columnwidth\hsize\csname@twocolumnfalse\endcsname
\title{Modeling of Spiking-Bursting Neural Behavior Using Two-Dimensional Map}
\author{Nikolai F. Rulkov}
\address{
Institute for Nonlinear Science, University of California, San
Diego, La Jolla, CA 92093-0402\\}

\date{\today}
\maketitle
\begin{abstract}
A simple model that replicates the dynamics of spiking and
spiking-bursting activity of real biological neurons is proposed.
The model is a two-dimensional map which contains one fast and one
slow variable. The mechanisms behind generation of spikes, bursts
of spikes, and restructuring of the map behavior are explained
using phase portrait analysis. The dynamics of two coupled maps
which model the behavior of two electrically coupled neurons is
discussed. Synchronization regimes for spiking and bursting
activity of these maps are studied as a function of coupling
strength. It is demonstrated that the results of this model are in
agreement with the synchronization of chaotic spiking-bursting
behavior experimentally found in real biological neurons.

\end{abstract}

\narrowtext
\vskip1pc]
\newcounter{eq}
\section{Introduction}
Understanding dynamical principles and mechanisms behind the
control of activity, signal and information processing that occur
in neurobiological networks is hardly possible without numerical
studies of collective dynamics of large networks of neurons. These
simulations need to take into account the complex architecture of
couplings among individual neurons that is suggested by data from
biological experiments. One of the complicating factors in
understanding the simulation results is the complexity of temporal
behavior of individual biological neurons. This complexity is due
to the large number of ionic currents involved in the nonlinear
dynamics of neurons. As a result, realistic channel-based models
proposed for a single neuron is usually a system of many nonlinear
differential equations (see, for example
\cite{HH52,Chay85,Chay90,Golovasch92,Gollomb93} and the review of
the models in~\cite{abarreview96}). The strong nonlinearity and
high dimensionality of the phase space is a significant obstacle in
understanding the collective behavior of such dynamical systems.
The dynamical mechanisms responsible for restructuring collective
behavior of a network of channel-based neuron models are difficult
or impossible to analyze because these mechanisms are well hidden
behind the complexity of the equations. If there is a chance to
identify possible dynamical mechanisms behind a particular behavior
of the network without the use of complex models, then this
knowledge can be used to guide through numerical study of this
behavior with the high-dimensional channel-based models. Based on
these results one can specify the actual dynamical mechanism
occurred in the network and understand the biological processes
which contribute to it.

One way to reveal the dynamical mechanisms is to use a simplified
or phenomenological model of the neuron. However, many experiments
indicate the restructuring of collective behavior utilizes a
variety of dynamical regimes generated by individual dynamics of a
neuron. When neural dynamics include the regime chaotic
spiking-bursting oscillations the system of differential equations
describing each neuron should be at least a three-dimensional
system (see for example~\cite{HRModel}). Further simplification of
phenomenological models for complex dynamics of the neuron can be
obtain using dynamical systems in the form of a map. Despite the
low-dimensional phase space of this nonlinear map, it is able to
demonstrate a large variety of complex dynamical regimes.

The use of low-dimensional model maps can be useful for
understanding the dynamical mechanisms if they mimic the dynamics
of oscillations observed in real neurons, show correct
restructuring of collective behavior, and are simple enough to
study the reasons behind such restructuring. The development of a
model map which is capable of describing various types of neural
activity, including generation of tonic spikes, irregular spiking,
and both regular and irregular bursts of spikes, is the goal of
this paper.

It is known that constructing a low-dimensional system of
differential equation which is capable of generating fast spikes
bursts excited on top of the slow oscillations, one needs to
consider a system which has both slow and fast dynamics (see for
example~\cite{HRModel,Rinzel85,Rinzel87,Belykh00,Izhikevich00}).
Using the same approach one can construct a two-dimensional map,
which can be written in the form
\setcounter{eq}{\value{equation}}
\addtocounter{eq}{+1}
\setcounter{equation}{0}
\renewcommand{\theequation}{\theeq \alph{equation}}

\begin{eqnarray}
 x_{n+1}&=&f(x_n,y_n+\beta_n)~, \label{mapx} \\
 y_{n+1}&=&y_n-\mu (x_n+1) + \mu \sigma_n~, \label{mapy}
\end{eqnarray}
\newcommand{\refmap}{1}
where $x_n$ is the fast and $y_n$ is the slow dynamical variable.
Slow time evolution of $y_n$ is due to small values of the
parameter $\mu=0.001$. Terms $\beta_n$ and $\sigma_n$ describe
external influences applied to the map. These terms model the
dynamics of the neuron under the action of the external DC bias
current $I_{DC}$ and synaptic inputs $I^{syn}_n$. The term
$\sigma_n$ can also be used as the control parameter to select the
regime of individual behavior.

\begin{figure}
\begin{center}
\leavevmode
\hbox{%
\epsfxsize=7cm
\epsffile{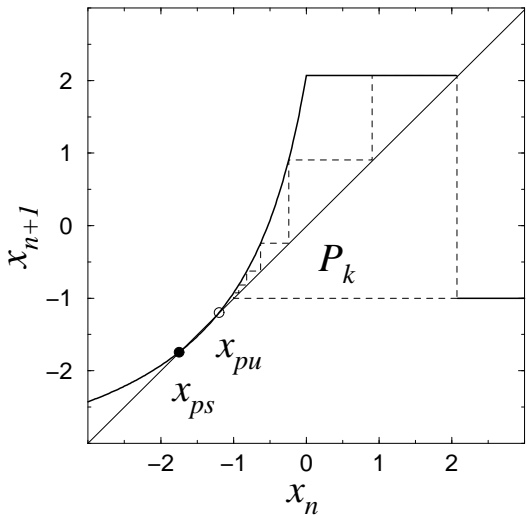}}
\end{center}
\caption{The shape of the nonlinear function $f(x,y)$, plotted for $\alpha=6.0$ and
$y=-3.93$, is shown with a solid line. The dashed line illustrates
a super-stable cycle $P_k$ of the fast map (\ref{umapx}), where the
value of $y$ is fixed. The stable and unstable fixed points of the
map are indicated by $x_{ps}$ and $x_{pu}$, respectively.}
\label{fig1}
\end{figure}

A model in the form of two-dimensional map, whose form is similar
to (\refmap) was used in the study of dynamical mechanisms behind
the emergence and regularization of chaotic bursts in a group of
synchronously bursting cells coupled through a mean
field\cite{rul_prl2001,deVries01}. The effect of anti-phase
regularization was also modeled before with one-dimensional
maps~\cite{Cazelles01}. In both these models the oscillations
during the burst were described by a chaotic trajectory. Map
(\refmap) improves these models by adding a feature that enables
one to mimic the dynamics of individual spikes within the burst.
This is achieved using a modification of the shape of the nonlinear
function $f(x,y)$ which is now a discontinuous function of the form
\setcounter{equation}{\value{eq}}
\renewcommand{\theequation}{\arabic{equation}}
\begin{eqnarray}
f(x,y)=\cases {
 \alpha/(1-x)+y,    & $x \leq 0$ \cr
 \alpha+y,          & $0<x<\alpha+y$ \cr
 -1,              & $x \geq \alpha+y$ \cr
            }
 \label{func}
\end{eqnarray}
where $\alpha$ is a control parameter of the map. The dependence of
$f(x,y)$ on $x$ computed for a fixed value of $y$ is shown in
Fig.\ref{fig1}. In this plot the values of $\alpha$ and $y$ are set
to illustrate the possibility of coexistence of limit cycle, $P_k$,
corresponding to spiking oscillation in (\ref{mapx}), and fixed
points $x_{ps}$ and $x_{pu}$. Note that when $y$ increases or
decreases the graph of $f(x,y)$ moves up or down, respectively,
except for the third interval $x\geq \alpha+y$, where the values of
$f(x,y)$ always remain equal to -1.

The paper is organized as follows: Section~\ref{Section1} considers
the features of the fast and slow dynamics which explain the
formation of various types of behavior in the isolated map. It also
discusses the bifurcations responsible for the qualitative change
of activity. The dynamics of response generated by the map after it
is excited or inhibited by an external pulse is discussed in
Section~\ref{Section2}. The goal of this section is to illustrate
dynamical mechanisms that control the response, and to show how the
relation between the two inputs of the map influence the properties
of the response. Section~\ref{Section3} presents the results of a
synchronization study in two coupled maps.

\section{Individual Dynamics of the Map}\label{Section1}

Consider the regimes of oscillations produced by the individual
dynamics of the map. In this case the inputs $\beta_n=\beta$ and
$\sigma_n=\sigma$ are constants. Note that if $\beta$ is a
constant, then it can be omitted from the equations using the
change of variable $y_n+\beta\rightarrow y_n^{new}$. Therefore, the
individual dynamics of the map depends only on the control
parameters $\alpha$ and $\sigma$, and the map can be rewritten in
the form
\setcounter{eq}{\value{equation}}
\addtocounter{eq}{+1}
\setcounter{equation}{0}
\renewcommand{\theequation}{\theeq \alph{equation}}
\begin{eqnarray}
x_{n+1}&=&f(x_n,y_n),
\label{umapx} \\
y_{n+1}&=&y_n-\mu (x_n+1) + \mu \sigma~. \label{umapy}
\end{eqnarray}
\newcommand{\refumap}{3}
Typical regimes of temporal behavior of the map are shown in
Figs~\ref{fig2} and \ref{fig3}. When the value of $\alpha$ is less
then 4.0 then, depending on the value of parameter $\sigma$, the
map generates spikes or stays in a steady state (see
Fig~\ref{fig2}). The frequency of the spikes increases as the value
of parameter $\sigma$ is increased (see Fig~\ref{fig2}).

\begin{figure}
\begin{center}
\leavevmode
\hbox{%
\epsfxsize=7.5cm
\epsffile{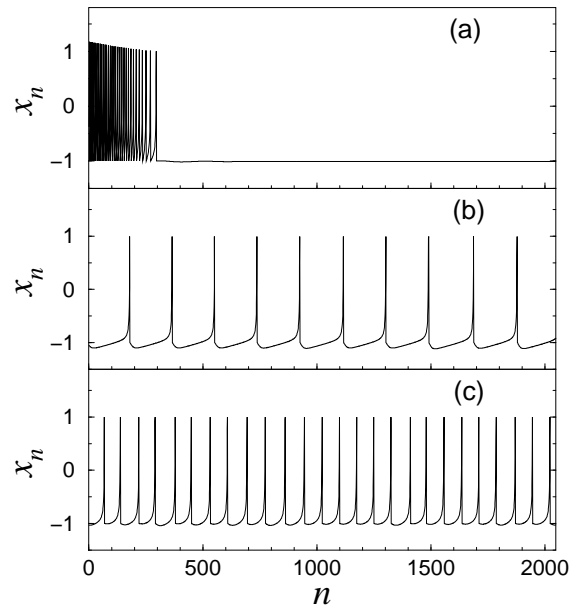}}
\end{center}
\caption{Waveforms of spiking behavior generated by map (\refumap)
with $\alpha=4$. Panel (a) shows the transition to
the regime of silence for $\sigma=-0.01$. The regimes of continuous
tonic spiking are computed for $\sigma=0.01$  (b) and $\sigma=0.1$
 (c).}
\label{fig2}
\end{figure}

For $\alpha>4$ the map dynamics are capable of producing bursts of
spikes. The spiking-bursting regimes are found in the intermediate
region of parameter $\sigma$ between the regimes of continuous
tonic spiking and steady state (silence). The spiking-bursting
regimes include both periodic and chaotic bursting. A few typical
bursting-spiking regimes computed for different values of $\sigma$
are presented in Fig~\ref{fig3}.

Due to the two different time scales involved in the dynamics of
the model, the mechanisms behind the restructuring of the dynamical
behavior can be understood through the analysis of the fast and
slow dynamics separately. In this case, time evolution of the fast
variable, $x$, is studied with the one-dimensional map
(\ref{umapx}) where the slow variable, $y$, is treated as a control
parameter whose value drifts slowly in accordance with equation
(\ref{umapy}). It follows from (\ref{umapy}) that the value of $y$
remains unchanged only if $x=x_{s}$ given by
\setcounter{equation}{\value{eq}}
\renewcommand{\theequation}{\arabic{equation}}
\begin{eqnarray}
x_{s}=-1 +\sigma.\label{xys}
\end{eqnarray}
If $x<x_{s}$, then the value of $y$ slowly increases. If $x>x_{s}$,
then $y$ decreases.

\begin{figure}
\begin{center}
\leavevmode
\hbox{%
\epsfxsize=7.5cm
\epsffile{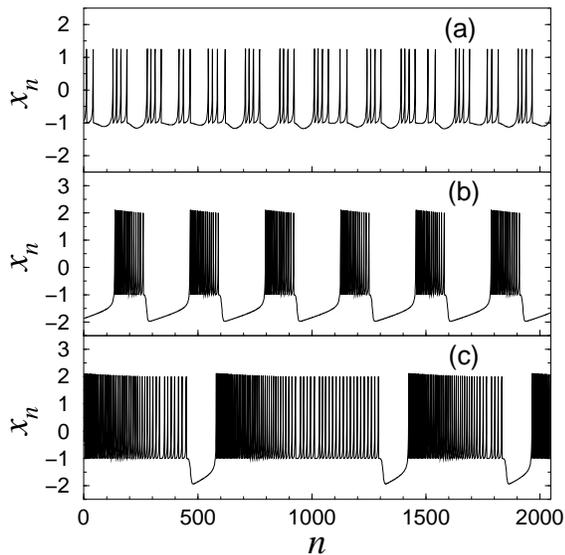}}
\end{center}
\caption{Typical waveforms of the spiking-bursting behavior generated
by the map (\refumap) for the following parameter values:
$\alpha=4.5$, $\sigma=0.14$  (a); $\alpha=6.0$, $\sigma=-0.1$  (b);
and $\alpha=6.0$, $\sigma=0.386$  (c).}
\label{fig3}
\end{figure}

\begin{figure}
\begin{center}
\leavevmode
\hbox{%
\epsfxsize=7cm
\epsffile{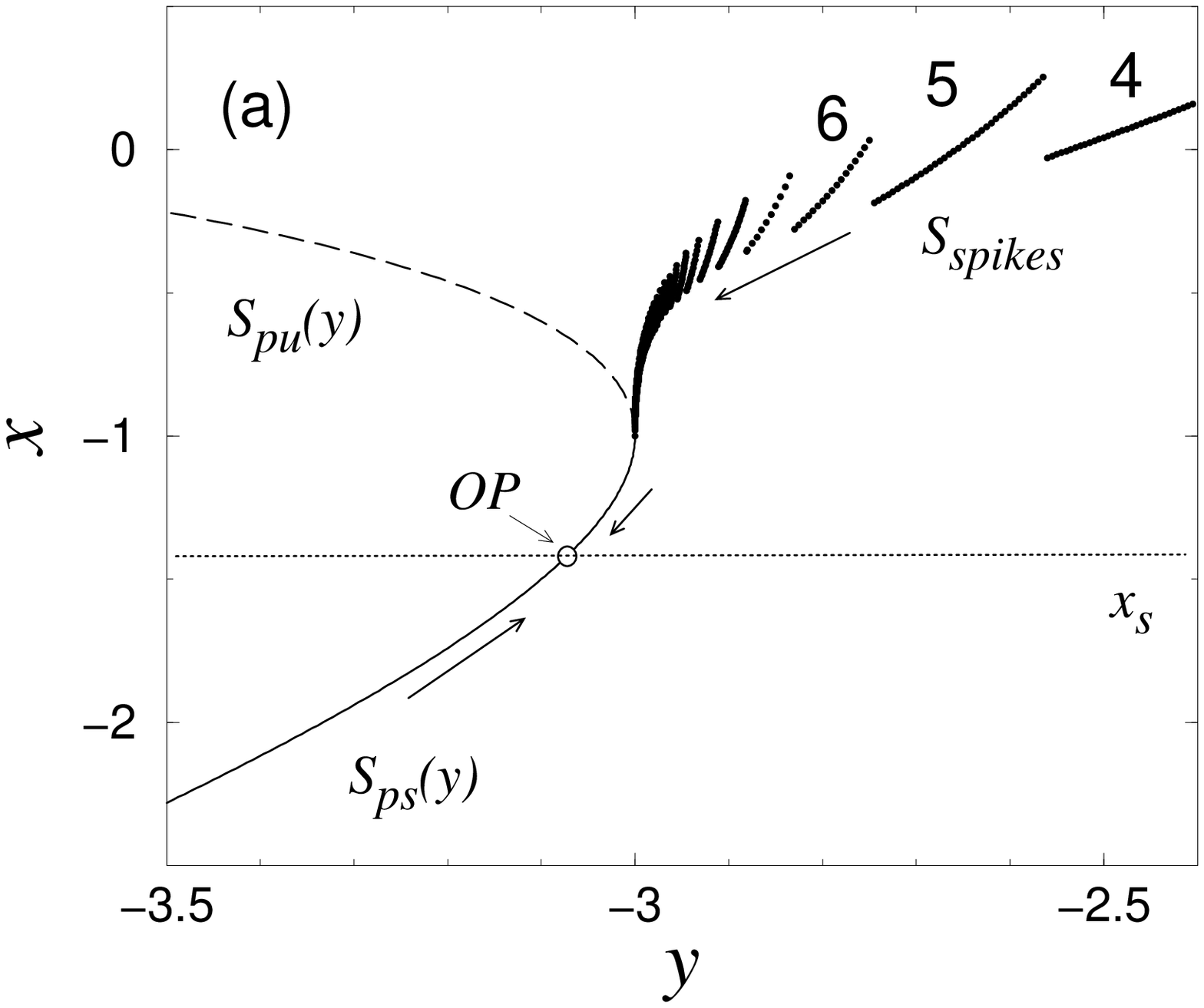}}
\hbox{%
\epsfxsize=7cm
\epsffile{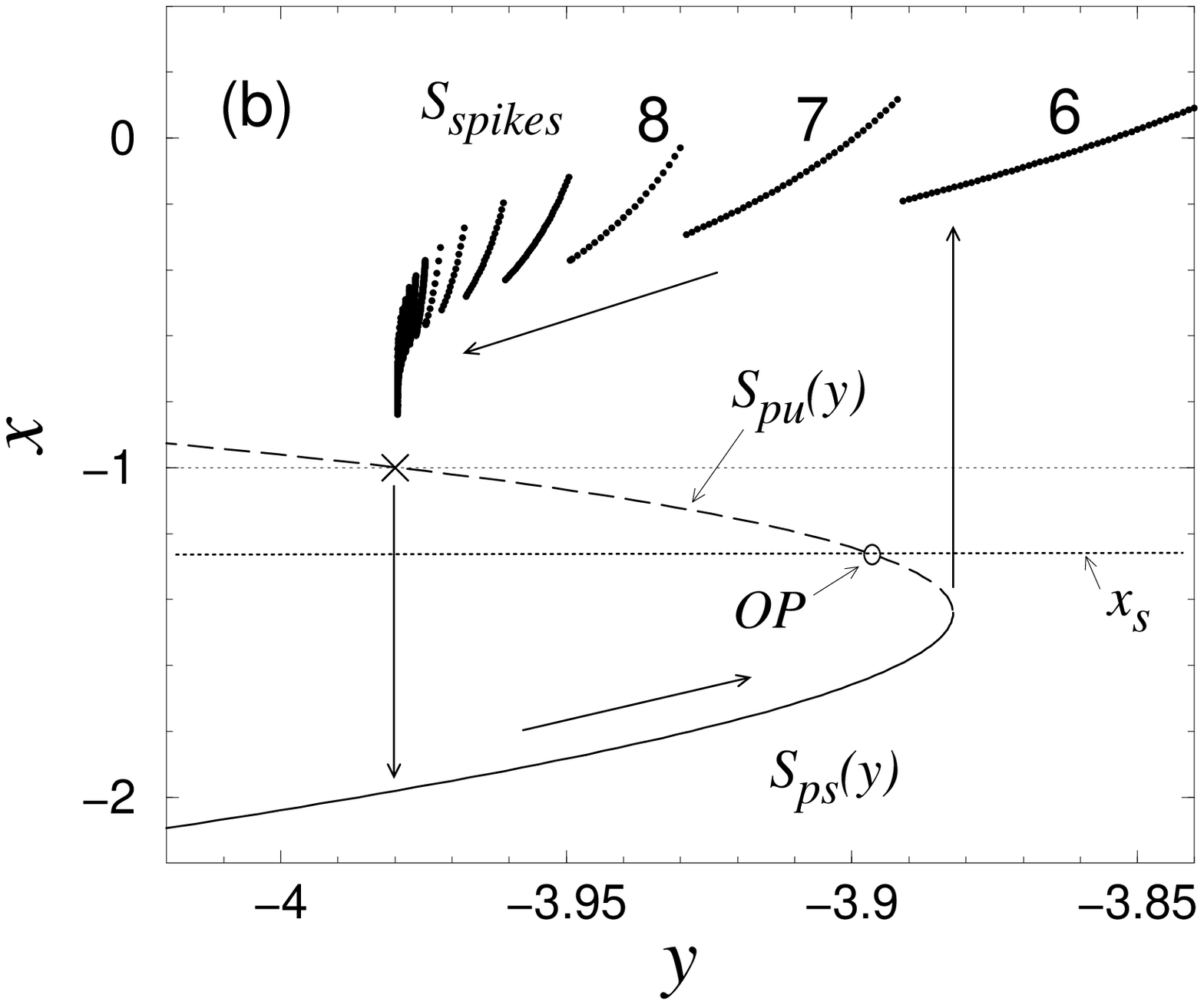}}
\end{center}
\caption{Stable ($S_{ps}(y)$, $S_{spikes}$) and unstable ($S_{pu}(y)$)
branches of slow dynamics of (\refumap) plotted on the plane of
phase variables ($y,x$). The cases of $\alpha=4$ and $\alpha=6$ are
presented in (a) and (b), respectively. Numbers on the branch
$S_{spikes}$ stand for the value of $k$ of the cycles $P_k$. The
operating points $OP$ are selected to illustrate the regime of
silence in (a) and the regime of spiking-bursting oscillations in
(b). Arrows indicate the direction of slow evolution along the
branches, and switching in the case of (b).}
\label{fig4}
\end{figure}

From (\ref{umapx}) one can find the equation for the coordinate of
fixed points $x_p$ of the fast map. This equation is of form
\begin{eqnarray}
y=x_p-\frac{\alpha}{1-x_p},\label{yvsx}
\end{eqnarray}
where $x_p \leq 0$. Equation (\ref{yvsx}) defines the branches of
slow motion in the two-dimensional phase space $(x_n,y_n)$ (see
Fig~\ref{fig4}).  The stable branch $S_{ps}(y)$ exists for
$x_p<1-\sqrt{\alpha}$ and the unstable branch $S_{pu}(y)$ exists
within $1-\sqrt{\alpha}\leq x_p \leq 0$

Considering the fast and slow dynamics together, one can see that,
if $x_s$ is in the stable branch $S_{ps}(y)$, then the map
(\refumap) has a stable fixed point. The stable fixed point
corresponds to the regime of silence in the neural dynamics. The
oscillations in the map dynamics will appear when
$x_s>1-\sqrt{\alpha}$. This is the threshold of excitation, which
corresponds to the bifurcation values of $\sigma$ given by
\begin{eqnarray}
\sigma_{th}=2-\sqrt{\alpha}.\label{sth}
\end{eqnarray}

To understand the dynamics of excited neurons within this model one
needs to consider branches that correspond to the spiking regime.
To evaluate the location of the spiking branches, consider the mean
value of $x_n$ computed for the periodic trajectory of the fast map
(\ref{umapx}) as a function of $y$. Here $y$ is treated as a
parameter. Use of such an approximation is quite typical for an
analysis involving fast and slow dynamics. It works well for small
values of parameter $\mu$.

It follows from the shape of $f(x,y)$ that for any value of $y$,
map (\ref{umapx}) generates no more than one periodic trajectory
$P_k$, where $k$ is the period (i.e. $x_n=x_{n+k}$). The cycles
$P_k$ always contain the point $x_n=-1$ and the index $k$ increases
stepwise ($k \rightarrow k+1$) as $y$ decreases. Since the
trajectory $P_k$ always has a point in the flat interval of
$f(x,y)$, all these cycles are super-stable except for bifurcation
values of $y$ for which the trajectory contains the point $x_n=0$
(see Fig~\ref{fig1}). The location of the spiking branch,
$S_{spikes}$, of "slow" motion in the phase plane ($x_n,y_n$) can
be estimated as the mean value of $x$ computed for the period of
cycle $P_k$,
\begin{eqnarray}
x_{mean}= \frac{1}{k} \sum_{m=1}^{k} f^{(m)}(-1,y), \label{xmean}
\end{eqnarray}
where $k$ is the period of $P_k$, and $f^{(m)}(x,y)$ is the $m$-th
iterate of (\ref{umapx}), started at point $x$ and computed for
fixed $y$. The spiking branch of "slow" dynamics evaluated with
(\ref{xmean}) is shown in Fig\ref{fig4}. One can see that this
branch has many discontinuities caused by the bifurcations of the
super-stable cycles $P_k$.

To complete the picture of fast and slow dynamics of the model for
$\alpha>4$, one needs to consider the fast map bifurcation
associated with the formation of homoclinic orbit $h_{pu}$
originating from the unstable fixed point, $x_{pu}$. This
homoclinic orbit occurs when the coordinate of $x_{pu}$ become
equal to -1. It can be easily shown that such a situation can take
place only if $\alpha>4$. The homoclinic orbit forms at the value
of $y$ where the unstable branch $S_{pu}(y)$ crosses the line
$x=-1$, see Fig.\ref{fig4}b. When the map is firing spikes and the
value of $y$ gets to the bifurcation point, the cycle $P_k$ merges
into the homoclinic orbit, disappears, and then the trajectory of
the map jumps to the stable fixed point $x_{ps}$.

Typical phase portraits of the model, obtained under the
assumptions made above, are presented in Fig.\ref{fig4}.
Fig.\ref{fig4}a shows the typical behavior for $2<\alpha\leq4$.
Here, only two regimes are generated. The first regime is the state
of silence, when the operating point (OP), given by the
intersection of $x_s=-1+\sigma$ and one of the branches, is on the
branch $S_{ps}(y)$. The second regime is the regime of tonic
spiking, when the operating point is selected on the spiking branch
$S_{spikes}$.

When $\alpha>4$ the picture changes qualitatively (see
Fig~\ref{fig4}b). Now, stable branches $S_{ps}(y)$ and $S_{spikes}$
are separated by the unstable branch $S_{pu}(y)$. If the operating
point is selected on $S_{pu}(y)$, then the phase of silence,
corresponding to slow motion along $S_{ps}(y)$, and spiking, when
system moves along $S_{spikes}$, alternate forming the regime of
spiking-bursting oscillations. The beginning of a burst of spikes
corresponds to the bifurcation state of the fast map where fixed
points $x_{ps}$ and $x_{pu}$ merge together and disappear. Before
this bifurcation the system is in $x_{ps}$ and, therefore, $y$
increases. The termination of the burst is due to the bifurcation
of the fast map associated with the formation of the homoclinic
orbit $h_{pu}$. Here, the limit cycle of the spiking mode merges
into the homoclinic orbit and disappears. After that the fast
subsystem flips to the stable fixed point $x_{ps}$. Then the
process repeats (see arrows in Fig~\ref{fig4}b).

It is clear from Fig.\ref{fig4}a that when operating point is set
on $S_{ps}(y)$ the model will be in the regime of silence. When the
operating point is set on the branch $S_{spikes}$ the model
produces tonic spiking, unless the point is set close to the
formation of homoclinic orbit $h_{pu}$. One can see that, at the
vicinity of this bifurcation, the branch $S_{spikes}$ becomes
densely folded. As a result, the behavior of $y$, which is governed
by the mean value of $x_n$, can become extremely sensitive to small
perturbations and even lead to instability caused by high-gain
feedback. This is one of the reasons for the irregular, chaotic
spiking-bursting behavior which occurs in the map when the
operating point is set close to the area of the branch $S_{spikes}$
where this branch is densely folded. The detailed and rigorous
analysis of chaotic dynamics cannot be done within the
approximations made above and require more precise computation of
$S_{spikes}$ which is beyond the scope of this paper.

\begin{figure}
\begin{center}
\leavevmode
\hbox{%
\epsfxsize=7cm
\epsffile{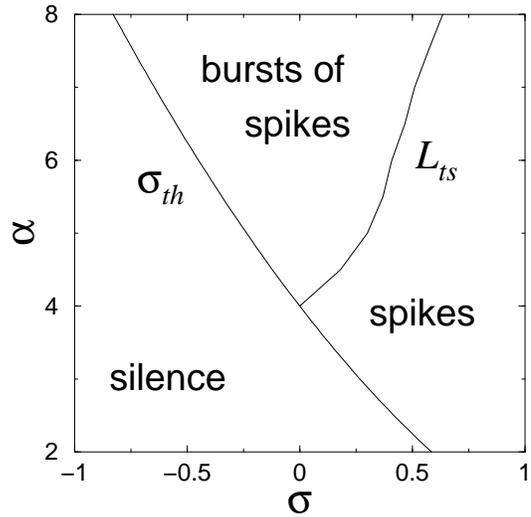}}
\end{center}
\caption{Bifurcation diagram on the parameter plane ($\sigma,\alpha$).}
\label{fig5}
\end{figure}

The results of the analysis presented above are summarized in the
sketch of the bifurcation diagram plotted on the parameter plain
($\sigma,\alpha$) (see Fig~\ref{fig5}). The bifurcation curve
$\sigma_{th}$ corresponds to the excitation threshold (\ref{sth})
where the fixed point of the 2-d map becomes unstable and the map
starts generating spikes. Curve $L_{ts}$ shows the approximate
location of the border between spiking and spiking-bursting
regimes, obtained in the numerical simulations. Note that
separation of spiking and bursting regimes is not always obvious,
especially in the regime of chaotic spiking. The regime of
spiking-bursting oscillations takes place within the upper triangle
formed by curves $\sigma_{th}$ and $L_{ts}$. The regimes of chaotic
spiking or spiking-bursting behavior are found in the relatively
narrow region of the parameters located around $L_{ts}$. This
region contains complex structure of bifurcations, associated with
multi-stable regimes, and is not shown in Fig~\ref{fig5}.

The bifurcation diagram shows the role of control parameters in the
selection of dynamical features of the considered neuron model.
Both parameters can be used to mimic a particular type of neural
behavior. Parameter $\sigma$ can be used to model the external DC
current injection that depolarizes or hyperpolarizes the neuron. In
this case $\sigma$ can be written as
\begin{eqnarray}
\sigma=\sigma_u+I_{DC}, \label{sig_dc}
\end{eqnarray}
where $\sigma_u$ is the parameter that selects the dynamics of
isolated neuron, and $I_{DC}$ is the parameter that models the DC
current injected into the cell. The changes in behavior caused by
$I_{DC}$ are similar to the action of the parameter $I$ in the well
known Hindmarsh-Rose model~\cite{HRModel}.

If the individual dynamics of the modelled neuron are capable of
generating only regimes of silence and tonic spiking, and do not
support the regime of bursts of spikes, then the value of $\alpha$
should be set below 4. In this case, no matter what the level of
$I_{DC}$ injected is, the system will not show bursts of spikes
(see Fig~\ref{fig5}).

\section{Modeling of Response to the Injection of Current}
\label{Section2}

To study the dynamical regimes of neuronal behavior in experiments,
biologists change the type of neural activity using the injection
of electrical current into the cell though an electrode. It was
shown above that the injection of DC current can be modeled in the
map (\refmap) using parameter $\sigma_n=\sigma$ (see equation
(\ref{sig_dc})). In this case, since the external influence does
not vary in time, the role of parameter $\beta_n=\beta$ is not
important, because the behavior of the map after the transient is
independent of the value of $\beta$. However, when the injected
current changes in time, it may be useful to consider the dynamics
of parameter $\beta_n$ in order to provide more realistic modeled
behavior during the transient. Taking this into account, the input
of the model can be considered in the form:
\begin{eqnarray}
\beta_n=\beta^e I_n~,~~ \sigma_n=\sigma^e I_n, \label{e_coupl}
\end{eqnarray}
where $I_n$ is injected current, and coefficients $\beta^e$ and
$\sigma^e$ are selected to achieve the desired properties of
response behavior.

This Section briefly illustrates how the relation between these
coefficients effects the dynamics of response to the pulse of
$I_n$. To be specific, consider the map in the regime of tonic
spiking with $\alpha=5.0$, $\sigma=0.33$, $\beta=0$. To study the
response behavior, positive and negative pulses of amplitude 0.8
and duration of 100 iterations were applied to the continuously
spiking map. In the simulations presented below the coefficient
$\sigma^e$ was selected to be equal to one.

\begin{figure}
\begin{center}
\leavevmode
\hbox{%
\epsfxsize=7cm
\epsffile{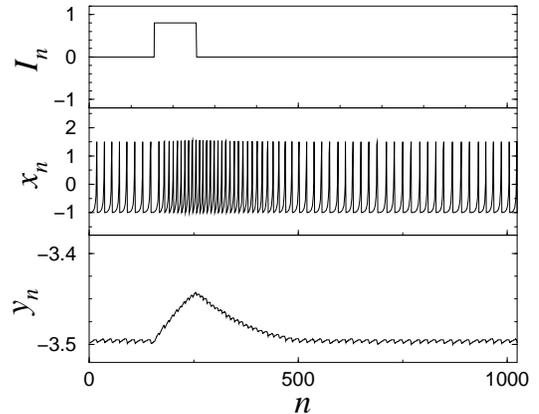}}
\end{center}
\caption{Response of the model with $\sigma^e=1$ and $\beta^e=0$
to a positive pulse of $I_n$. The parameters of the map are
selected in the regime of tonic spiking ($\alpha=5.0$,
$\sigma=0.33$, $\beta=0$).}
\label{fig6}
\end{figure}

\begin{figure}
\begin{center}
\leavevmode
\hbox{%
\epsfxsize=7cm
\epsffile{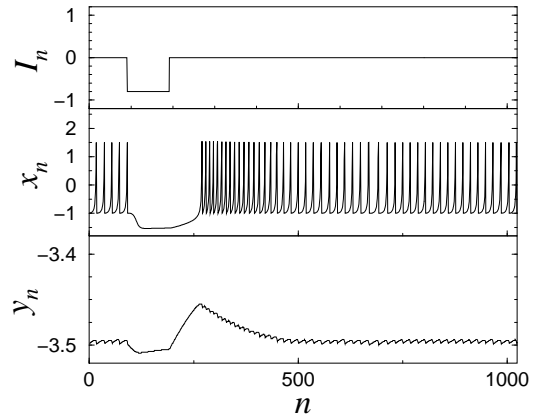}}
\end{center}
\caption{Response of the model with $\sigma^e=1.0$ and $\beta^e=0$ to
a negative pulse of $I_n$. The parameters of the map are selected
in the regime of tonic spiking ($\alpha=5.0$, $\sigma=0.33$,
$\beta=0$).}
\label{fig7}
\end{figure}

Figures~\ref{fig6} and \ref{fig7} show the response to the positive
and negative pulse, respectively, computed with $\beta^e=0$. In
this case, during the action of a positive pulse, the value of
$y_n$ increases monotonically because of the increased value of
$\sigma_n$ (see (\ref{mapy})). The increase of $y_n$ pushes the
fast map up, see Fig~\ref{fig1}. This leads to  an increase in the
frequency of spiking. After the action of the pulse ends the value
of $y_n$ monotonically decreases back to the original state (see
Fig.~\ref{fig6}).

When a negative pulse is applied, it pushes the operating point
down and, if the amplitude of the pulse is sufficiently large,
shuts off the regime of spiking (see Fig.\ref{fig7}). This happens
because the trajectory of the fast map gets to the perturbed
operating point which is now on the stable branch $S_{ps}(y)$, (see
Fig~\ref{fig4}b). After the pulse is over the spiking does not
re-appear immediately because the system spends some time drifting
along the stable branch of slow motions $S_{ps}(y)$, during which
the variable $y$ overshoots its original value for the spiking
regime. As a result, after the system switches to the spiking
branch $S_{spikes}$, $y_n$ monotonically drifts down to the
unperturbed operating point. The dynamics of the slow evolution are
clearly seen in the lower panel of Fig.\ref{fig7}.

\begin{figure}
\begin{center}
\leavevmode
\hbox{%
\epsfxsize=7cm
\epsffile{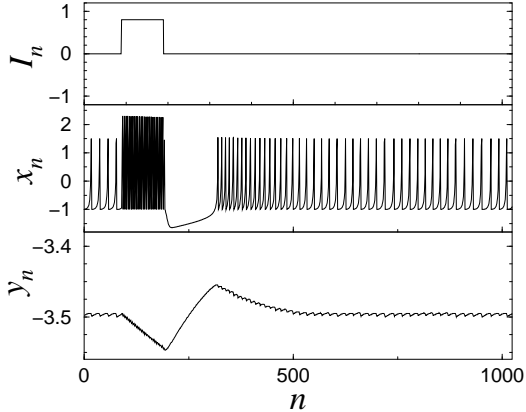}}
\end{center}
\caption{Response of the model with $\sigma^e=1.0$ and $\beta^e=1.0$
to a positive pulse of $I_n$. The parameters of the map are
selected in the regime of tonic spiking ($\alpha=5.0$,
$\sigma=0.33$, $\beta=0$).}
\label{fig8}
\end{figure}

\begin{figure}
\begin{center}
\leavevmode
\hbox{%
\epsfxsize=7cm
\epsffile{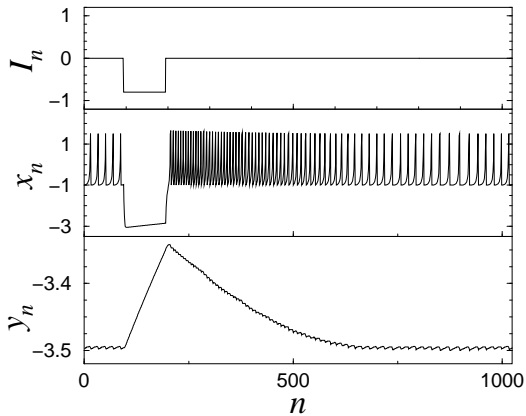}}
\end{center}
\caption{Response of the model with $\sigma^e=1.0$ and $\beta^e=1.0$
to a negative pulse of $I_n$. The parameters of the map are
selected in the regime of tonic spiking ($\alpha=5.0$,
$\sigma=0.33$, $\beta=0$).}
\label{fig9}
\end{figure}

Figures~\ref{fig8} and \ref{fig9} illustrate how response to the
pulse of $I_n$ changes when coefficient $\beta_n$ is not equal to
zero. To be brief and specific consider the case of $\beta^e=1.0$
and $\sigma^e=1.0$. Comparing Fig.\ref{fig8} with Fig.\ref{fig6}
one can see that the response dynamics to the positive pulse
changes qualitatively. Indeed, when the pulse current is applied,
then, acting through $\beta_n$, it immediately forces the fast map
to shift up. As a result, the rate of spiking and mean value of
$x_n$ increases sharply. Variable $y_n$ reacts to this change and
decreases its value to compensate for the sudden change of the mean
value of $x_n$. When the action of the pulse is over, the value of
$\beta_n$ returns to its original value and, due to the updated
levels of $y_n$, the fast map overshoots its original state. As a
result, the trajectory of the fast map reaches the stable fixed
point. To return to the original regime of spiking the system has
to go all the way along the branches of slow motion $S_{ps}(y)$ and
$S_{spikes}$ back to the original operating point (see
Fig.~\ref{fig4}b). This type of response is observed in real
neurobiological experiments, see for example~\cite{Kleinfeld90}.

Using the same analysis one can understand the new effects in the
response to a negative pulse caused by the action of $\beta_n$.
These new effects can be clearly seen comparing Fig.\ref{fig9} with
Fig.~\ref{fig7}.

The presented results illustrate how 2-d map (\refmap) along with
equation~(\ref{e_coupl}) can model a large variety of transient
neural behavior induced by injected current. Due to the simplicity
of the model one can clearly see the nonlinear mechanisms behind
the response behavior and apply them to select the desired balance
between $\sigma^e$ and $\beta^e$ to model a particular type of
response.

\section{Regimes of synchronization in two coupled maps}
\label{Section3}

This section presents the results of studies of synchronization
regimes in coupled chaotically bursting 2-d maps. The goal of this
study is to reproduce the main regimes of synchronous behavior
found in a real neurobiological experiment~\cite{elson98}. The
experiment was carried out on two electrically coupled neurons (the
pyloric dilators, PD) from the pyloric CPG of the lobster
stomatogastric ganglion~\cite{CPG}. The regimes found in the
experiment were also reproduced in numerical simulations using
Hindmarsh-Rose model~\cite{Abarbanel96,Pinto2000}, one-dimensional
map model~\cite{Cazelles01}, and in experiments with electronic
neurons~\cite{Pinto2000}.

The equations used in the numerical simulations of the coupled maps
are of form
\begin{eqnarray}
x_{i,n+1}&=&f(x_{i,n},y_{i,n}+\beta_{i,n}),\nonumber
\\
y_{i,n+1}&=&y_{i,n}-\mu (x_{i,n}+1) + \mu \sigma_i+\mu
\sigma_{i,n},
\label{twocells}
\end{eqnarray}
where index $i$ specifies the cell, and $\sigma_i$ is the parameter
that defines the dynamics of the uncoupled cell. The coupling
between the cells is provided by the current flowing from one cell
to the other. This coupling is modeled by
\begin{eqnarray}
\beta_{i,n}&=&g_{ji} \beta^e (x_{j,n}-x_{i,n}) \nonumber
\\
\sigma_{i,n}&=&g_{ji} \sigma^e (x_{j,n}-x_{i,n}) \label{twocoupl}
\end{eqnarray}
where $i \neq j$, and $g_{ji}$ is the parameter characterizing the
strength of the coupling. The coefficients $\beta^e$, $\sigma^e$
set the balance between the couplings for the fast and slow
processes in the cells, respectively. In the numerical simulations
the values of the coefficients are set to be equal: $\beta^e=1.0$
and $\sigma^e=1.0$. The other parameters of the coupled
maps~(\ref{twocells}) that remain unchanged in the simulations have
the following values: $\mu=0.001$, $\alpha_1=4.9$, $\alpha_2=5.0$.
The coupling between the maps is symmetrical, $g_{ji}=g_{ij}=g$.

\begin{figure}
\begin{center}
\leavevmode
\hbox{%
\epsfxsize=8.5cm
\epsffile{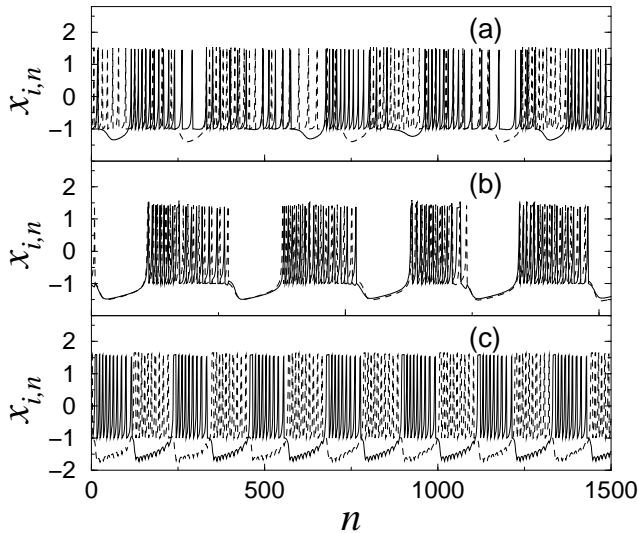}}
\end{center}
\caption{Waveforms generated by the two coupled 2-d maps (\ref{twocells})
with $\sigma_1=0.240$ and $\sigma_2=0.245$. Waveforms of $x_{1,n}$
and $x_{2,n}$ are shown by solid and dashed lines, respectively.
Three different regimes of spiking-bursting behavior are shown.
Individual chaotic spiking-bursting oscillations of uncoupled maps,
$g=0.0$, - (a). Regime of synchronized chaotic bursts, computed
with $g=0.043$, - (b). Regime of antiphase synchronization,
computed with $g=-0.029$, -(c).}
\label{fig10}
\end{figure}

First, consider the main regimes of synchronization between the
maps generating irregular, chaotic bursts of spikes. To set the
individual dynamics of the maps to this regime, the parameters
$\sigma_i$, that take into account the DC bias current injected
into the neurons (see~\cite{elson98} for details), are tuned to the
following values: $\sigma_1=0.240$, $\sigma_2=0.245$. When these
maps are uncoupled ($g=0$) they produce chaotic spiking-bursting
oscillations shown in Fig~\ref{fig10}a. When the coupling becomes
sufficiently large the slow components of the bursts synchronize
while spikes within the bursts remain asynchronous see the
waveforms in Fig.~\ref{fig10}b obtained with $g=0.043$. This regime
of synchronization is typical for naturally coupled PD neurons (see
Figure 2a in~\cite{elson98}).

Introduction of negative coupling, $g<0$, also leads to
synchronization of bursts, but in this regime of synchronization
the systems burst in antiphase. Typical waveforms produced in this
regime are presented in Fig.~\ref{fig10}c. This regime of
synchronization of chaotic bursts is also observed in the
experimental study of coupled PD neurons (see Figure 2c in
~\cite{elson98}). It is important to emphasize that, both in this
simulation and in the experiment, the regime of antiphase
synchronization characterized by the onset of regular bursts.

The simplicity of this model enables one to understand the possible
cause for the onset of regular bursting in the antiphase
synchronization. It is shown in Section~\ref{Section1} that chaotic
dynamics of bursts occurs when the operating point of the system
appears close to the leftmost area of the spiking branch
$S_{spikes}$. In this region, $S_{spikes}$ is densely folded and,
if system slows down in this area, the timing for the end of the
burst becomes very sensitive to infinitesimal perturbations.

\begin{figure}
\begin{center}
\leavevmode
\hbox{%
\epsfxsize=7cm
\epsffile{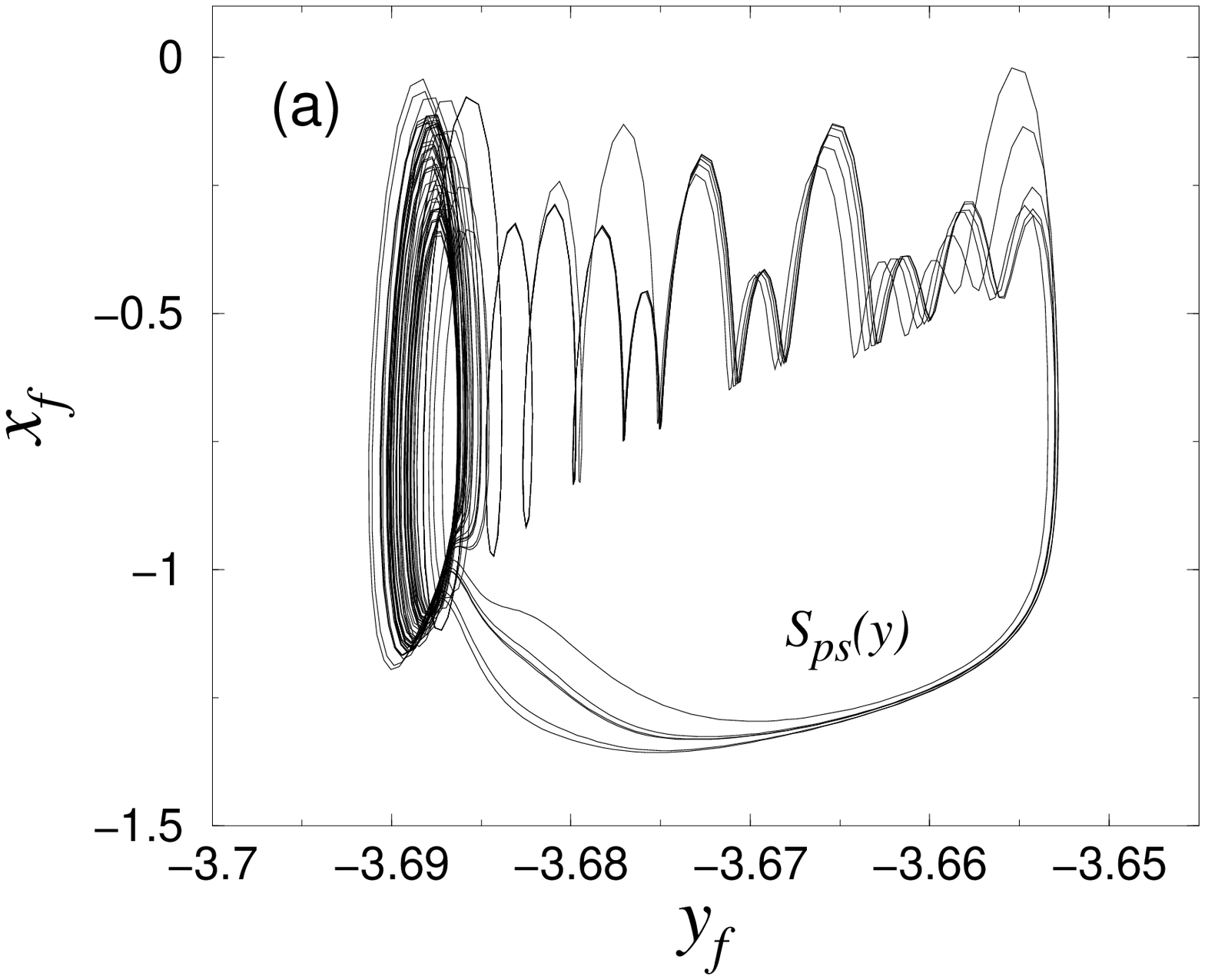}}
\hbox{%
\epsfxsize=7cm
\epsffile{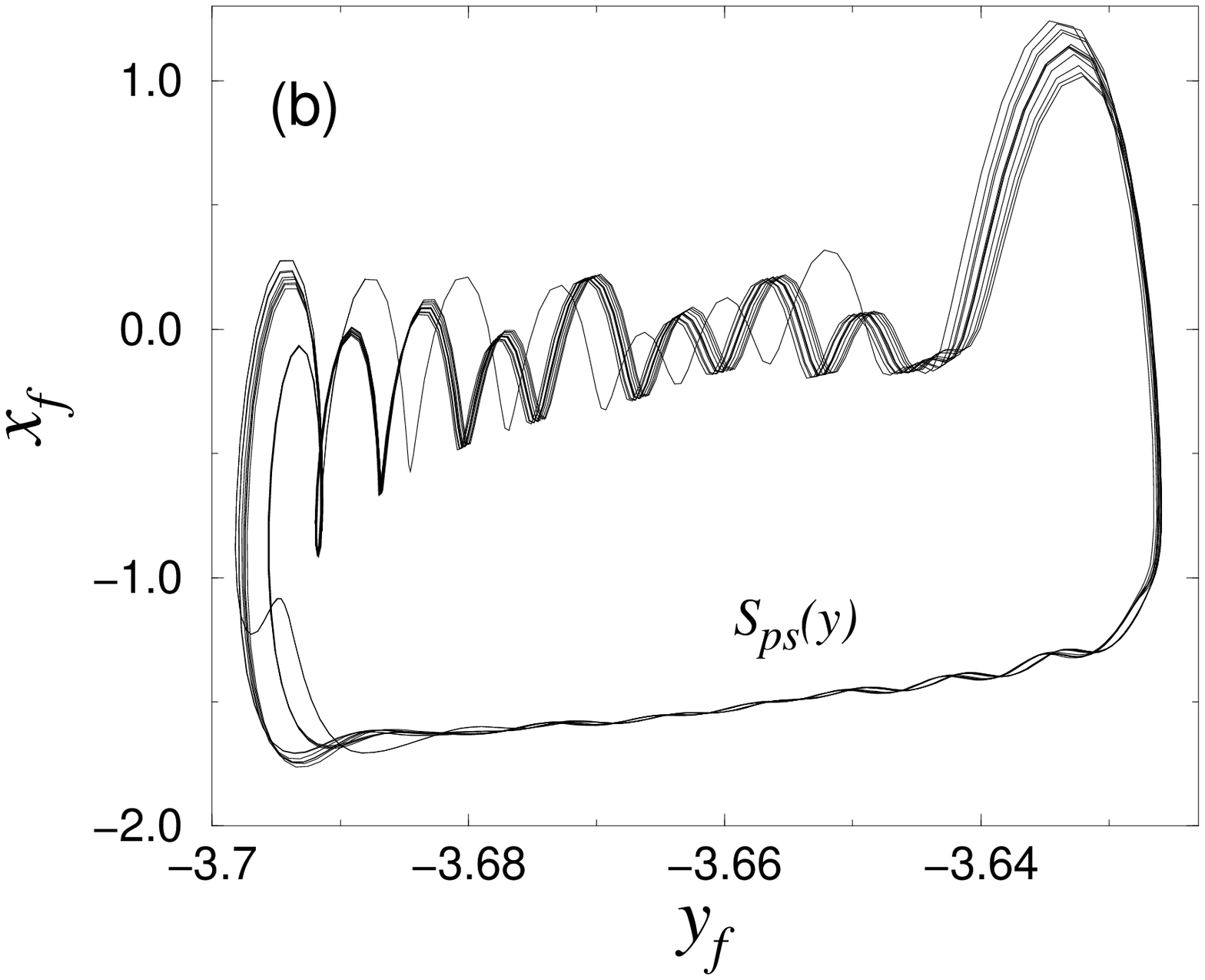}}
\end{center}
\caption{Attractors computed from the waveforms of the first cell
operating in the regime of uncoupled oscillations - (a) and in the
regime of antiphase synchronization - (b). Waveforms for these
regimes are shown in Fig.~\ref{fig10}a and Fig.~\ref{fig10}c,
respectively. The points of the attractors are connected by lines
to clearly show their sequence in time. }
\label{fig11}
\end{figure}

This fact is illustrated in Fig.~\ref{fig11}a. This figure shows
the attractor computed from the waveforms $x_{1,n}$ and $y_{1,n}$
of the first cell when the cells are uncoupled (this regime is
shown in Fig.~\ref{fig10}a). To plot the attractor, the waveforms
of $x_{1,n}$ and $y_{1,n}$ were filtered by a fourth order low-pass
filter with cutoff frequency 0.6. As a result, the attractor does
not contain sharp spikes and the approximate location $S_{spikes}$,
which is close to the center of filtered spiking oscillations, is
easy to see. The level of coordinate $x_f$, corresponding to the
operating point, is equal to $x_s=-1+\sigma_1=-0.76$. One can see
from Fig.~\ref{fig11}a that when the center of oscillations gets
close to that level the trajectory becomes rather complex, as shown
by the dense set of trajectories in the left part of the attractor.

The numerical simulations show that this effect of regularization
takes place even when $\sigma^e=0$. Therefore, the vertical shift
of operating point, $x_s$, caused by the slow coupling current is
not very important for this effect. The coupling term $\beta_{i,n}$
directly influences the fast dynamics by shifting the graph of the
1-d map (see Fig.~\ref{fig1}) up or down depending on the sign of
$\beta_{i,n}$. In the regime of antiphase synchronization
$\beta_{i,n}$ is positive, when the $i$-th cell is spiking, and
negative, when it is silent. Therefore, from the viewpoint of fast
dynamics, the coupling forces the cells to stay on the current
branch of slow motion. One can understand these dynamics from the
graph of $f(x,y)$ and (\ref{mapx}). This effect is also seen as the
formation of extended shape of attractor plotted for the regime of
antiphase synchronization (see Fig.~\ref{fig11}b). Note that axes
in Fig.~\ref{fig11}a and Fig.~\ref{fig11}b have different scales.

\begin{figure}
\begin{center}
\leavevmode
\hbox{%
\epsfxsize=8.5cm
\epsffile{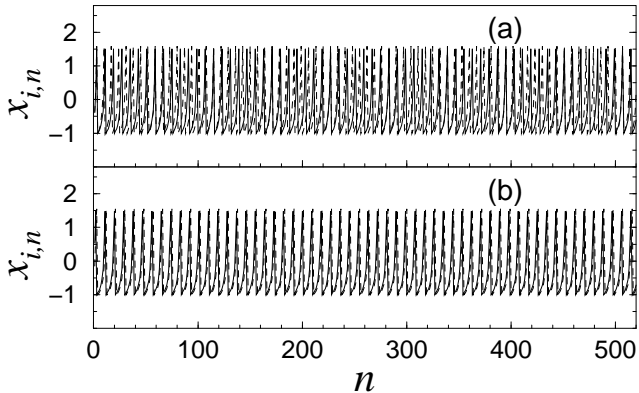}}
\end{center}
\caption{Tonic spiking waveforms generated with $\sigma_1=0.653$
and $\sigma_2=0.714$. Waveforms of $x_{1,n}$ and $x_{2,n}$ are
shown by solid and dashed lines, respectively. Spiking in the
uncoupled maps, $g=0.0$, - (a). Regime of synchronized spikes,
computed with $g=0.008$, - (b)}
\label{fig12}
\end{figure}

In the regime of chaotic bursts the level of $x_s$ is close to the
stable branch of spiking $S_{spikes}$, and, therefore, the slow
evolution along the branch of silence $S_{pu}$ is faster then along
$S_{spikes}$. This means that duration of the phase of silence in
the cell is shorter than duration of the burst of spikes. When the
cell switches from silence to spiking it sharply changes the value
and the sign of $\beta_{i,n}$. This change pushes $S_{spikes}$ of
the spiking cell down to its original location, and, as a result,
quickly drags the trajectory through the area of complex behavior
toward the branch $S_{pu}$. Due to this fast change the duration of
bursting is determined only by the dynamics of the silent phase,
which is regular.

It was observed in the experiment that, in the regime of tonic
spiking, spikes in PD neurons synchronize at low levels of coupling
(i.e. without added artificial coupling, see~\cite{elson98} for
details). This property of synchronization in the regime of tonic
spiking is also typical for the dynamics of the maps considered
here (see Fig.\ref{fig12}). In these numerical simulations, the
uncoupled maps were tuned to generate spikes with slightly
different rates. One can clearly see the beats between the
waveforms plotted in Fig.\ref{fig12}a. When small coupling,
$g=0.008$, is introduced the beats disappear as soon as the spikes
get synchronized (see Fig.~\ref{fig12}b).

Although the synchronization between continuously spiking maps is
easily achieved, it is important to understand that the dynamics of
such synchronization in the maps is a much more complex process
than it is in the models with continuous time. Due to the discrete
time, the periodic spikes can lock with different frequency ratios.
This results in the existence of a complex multistable structure of
synchronization regimes. This structure becomes more noticeable in
the dynamics of synchronization as the number of iterations in the
period of spikes decreases.

\section{Discussion and outlook}\label{Section4}

The simple phenomenological model of complex dynamics of
spiking-bursting neural activity is proposed. The model is given by
two-dimensional map (\refmap). The first equation (\ref{mapx}) of
the map describes the fast dynamics. Its isolated dynamics is
capable of generating stable limit cycles, which mimics the spiking
activity of a neuron, and a stable fixed point, which corresponds
to phase of silence. This 1-d map has a region of parameters were
both these stable regimes coexist. Existence of such multistable
regimes is the reason for generation of bursts, when the operating
point, defined by the dynamics of the second equation (\ref{mapy}),
is properly set (see Fig~\ref{fig4}b).

The shape of the nonlinear function $f(x,y)$ used in (\ref{mapx})
is selected in the form (\ref{func}) based on the following
considerations: (i) The fast map should generate limit cycles whose
waveforms mimics those of the spikes. (ii) Each spike generated by
the map always has a single iteration residing on the most right
interval $x\geq\alpha+y$. Therefore, the moment of time
corresponding to the appearance of a trajectory on this interval
can be used to define the time of a spike. This feature is
important for modeling the dynamics of chemical synapses in a group
of coupled neurons. (iii) The analytic expression of the nonlinear
function in interval $x<0$ should be simple enough to allow
rigorous analysis of bifurcations of fixed points of the fast map.
(iv) Use of the fixed level, $-1$, in the rightmost interval of the
function simplifies the analysis of dynamics at the end of the
bursts. The end of a burst is associated with the formation of a
homoclinic orbit, which corresponds to the case when the trajectory
from this interval maps into the unstable fixed point (see
Section~\ref{Section1}).

It is clear that the shape of function (\ref{func}) can be modified
to take into account other dynamical features that need to be
modeled. Due to the low-dimensionality of the model, the dynamical
mechanisms behind its behavior are easy to understand using phase
plane analysis. This allows one to see how introduced modifications
influence the dynamics.

Some modification of the fast map is required to enhance the region
of parameters where the model can still be used to mimic neural
dynamics properties. For example, from (\ref{func}) and
Fig.~\ref{fig1} one can see that, due to the dynamics of coupling
terms, the trajectory of the fast system can stay in the middle
interval of $f(x,y)$ for several iterations. This can happen when
the external influence monotonically pushes the function up while
the trajectory is located in the middle interval. In this case the
trajectory will map to the middle interval again and again,
increasing the duration of a spike. This artifact can be removed
using one of the following modifications. Introduction of a
sufficient gap between the right end of this interval and the
diagonal (see Fig.~\ref{fig1}) will help the map to terminate the
spike despite the monotonic elevation of the function.
Alternatively one can introduce an additional condition to the fast
map (\ref{mapx}) that forces the map always to iterate its
trajectory from the middle interval to the rightmost one, despite
the dynamics of $y$ (see (\ref{func})). This can be achieved using
the function $f(x_n,y)$ of the following form
\begin{eqnarray}
f(x_n,y)=\cases {
 \alpha/(1-x_n)+y,    & $x_n \leq 0$ \cr
 \alpha+y,          & $0<x_n<\alpha+y$ \cr
 -1,              & $x_n \geq \alpha+y$ or $x_{n-1}>0$\cr
            }
\nonumber
\end{eqnarray}

An important feature of the model discussed here is that one can
use two inputs, $\beta_n$ and $\sigma_n$, to achieve the desired
response dynamics. Although these inputs are not directly related
to dynamics of specific ionic currents, they can be used to capture
the collective dynamics of these currents. Selecting a proper
balance between these inputs, one can model a large variety of the
responses that are seen in different neurons. Again, the simplicity
of the model helps one to understand the dynamical properties of
each input and set the proper balance between them.

\section{Acknowledgment}

The author is grateful to M.I. Rabinovich, R. Elson, A. Selverston,
H.D.I. Abarbanel, P. Abbott, V.S. Afraimovich  and A.R. Volkovskii
for helpful discussions. This work was supported in part by U.S.
Department of Energy (grant DE-FG03-95ER14516), the U.S. Army
Research Office (MURI grant DAAG55-98-1-0269), and by a grant from
the University of California Institute for Mexico and the United
States (UC MEXUS) and the Consejo Nacional de Ciencia y Tecnologia
de M\'{e}xico (CONACYT).

\end{document}